%% file: chaudhuriHandcock.tex
\documentclass[11pt]{amsart}

\usepackage{natbib}
\usepackage{caption}
\usepackage{fullpage}

\usepackage{csquotes}

\usepackage{csquotes,amsmath,amssymb, amsbsy, mdwlist}

\newcounter{blind}
\newcommand{\blind}[2]{\ifthenelse{\equal{\value{blind}}{0}}{#1}{#2}}
\usepackage{ifthen}
\usepackage[usenames]{color}
\newcounter{content}
\newcounter{notes}
\newcounter{techreport}

\usepackage{amssymb,amsmath}
\usepackage{mathrsfs} 
\usepackage{amsmath}
\usepackage{amssymb}
\usepackage[all]{xy}
\usepackage{color, graphics}
\usepackage{subfigure}
\usepackage{mathrsfs}
\usepackage{multicol}
\usepackage{tabularx}

\definecolor{Emphcolor}{cmyk}{0,0.89,0.94,0.1}
\definecolor{Netcolor}{rgb}{.8,0,.9}
\definecolor{Diseasecolor}{rgb}{1,.8,.2}
\definecolor{Sampcolor}{rgb}{0,.9,.3}
\definecolor{Black}{rgb}{0,0,0}
\definecolor{Red}{rgb}{1,0,0}
\definecolor{Blue}{rgb}{0,0,1}
\definecolor{Gray}{gray}{.6}

\font\elevenrm=cmr11

\newcommand{\qrns}{{\elevenrm "}\kern-2pt}
\newcommand{\pkg}[1]{{\normalfont\fontseries{b}\selectfont #1}}
\newcommand{\proglang}[1]{{\textsf #1}}

\theoremstyle{definition}

\usepackage{float}
\usepackage{rotating}
\usepackage{color}
\usepackage{hypernat}
\usepackage{pdflscape}

\usepackage{graphics}
\usepackage{subfigure}
\usepackage[all]{xy}
\usepackage{amsmath}
\usepackage{amssymb}
\usepackage[all]{xy}
\usepackage{multicol}
\usepackage{enumerate}

\newcommand{\mc}[1]{\mathcal{#1}}
\newcommand{\mbb}[1]{\mathbb{#1}}
\newcommand{\h}[1]{\hat{#1}}

\newcommand{\mG}{\mathbb{G}}

\newcommand{\ind}{\perp\!\!\!\perp}
\newcommand{\cind}[3]{#1\ind#2\mid#3}
\newcommand{\bp}{\nu}
\newcommand{\Bp}{{\scriptstyle\Upsilon}}
\newcommand{\pf}{F_{\mc{S}}}
\newcommand{\argmax}{\arg~\max}
\newtheorem{assumption}{Assumption}
\newtheorem{lemma}{Lemma}[section]
\newtheorem{theorem}{Theorem}[section]
\newtheorem{corollary}{Corollary}[section]

\title[]{A Conditional Empirical Likelihood Based Method for Model Parameter Estimation from Complex survey Datasets}
\author{Sanjay Chaudhuri}
\address{Sanjay Chaudhuri, Department of Statistics and Applied Probability, National University of Singapore, Singapore $117546$.}
\email{sanjay@stat.nus.edu.sg}
\thanks{We acknowledge partial support for this work from AcRF grant R-155-000-176-114, the National Science Foundation (NSF, MMS-0851555, SES-1357619, IIS-1546259), National Institute of Child Health and Human Development (NICHD, R21HD063000, R21HD075714 and R24-HD041022).\\This article appears in \emph{Statistics and Applications}, Volume {\bf 16} No. 1, 2018 (New Series), pp 245-268.
}
\author{Mark S. Handcock}
\address{Mark S. Handcock, Department of Statistics, University of California, Los Angeles, CA $90095-1554$.}  \email{handcock@ucla.edu}

\begin{document}

\begin{abstract}
We consider an empirical likelihood framework for
inference for a statistical model 
based on an informative sampling design.
Covariate information is incorporated both through the weights and 
the estimating equations.
The estimator is based on conditional weights.
We show that under usual conditions, with population size increasing unbounded, the estimates are strongly consistent,
asymptotically unbiased and normally distributed.  
Our framework provides
additional justification for inverse probability weighted score
estimators in terms of conditional empirical likelihood.  
In doing so, it bridges the gap between design-based and model-based modes of inference in
survey sampling settings.
We illustrate these ideas with an application to an electoral survey.

\end{abstract}


\maketitle

\section{Introduction}
Because of their easy interpretability, parametric models are popular in statistics for explaining natural phenomenon.  These model parameters are usually estimated by maximising the so called likelihood function computed from preferably independent sample observations identically distributed according to the model in the population. 
In practice however, such data-sets are often unavailable.  More often than not, practitioners are forced to work with data obtained from various surveys. 
In real world, surveys are complex.  Observations are drawn according to
informative designs and are accompanied by unequal sampling weights.  Because of such complex sampling the observed data distribution varies from the distribution specified in the model.     
These sampling weights thus contain important information about the distribution in the population and cannot be ignored. In most situations, ignoring the weights leads to sevrely biased and/or inefficient estimators \citep{shs_1989}.

There is a large literature dealing with the analysis of complex survey data.  However, the incorporation of design weights in a parametric likelihood framework is difficult and many common approaches are not fully likelihood based.  In most cases design unbiased estimators for 
large but finite population parameters are used \citep{narain_1951,horvitz_thompson_1952} and the model parameters are assumed to be close to this estimator.  

Empirical likelihood \citep{owenbook} provides alternative to such design unbiased estimators.  For data sampled with equal probabilities, empirical likelihood based procedures re-weighs the data points with unknown weights. These weights are then estimated from the data by maximising their product under various constraints.  
These constraints can involve unknown parameters which can also be estimated simultaneously \citep{Qinlawless1}.  The constraints can also represent known characteristics of the population obtained from census, registration data etc. \citep{chaudhuri_handcock_rendall_2008}.  
Historically, the term was coined and made popular by \citet{Owen_1988}.  However, it has been argued that precursor methods were available \citep[see, for example,][]{hartley_rao_1968, thomas_grunkemeier_1975}.

Empirical likelihood based methods which take into account the
design weights in the sample have been studied by several
researchers.  \citet{chen_sitter_1999} were motivated by the
Horvitz-Thompson estimator in survey sampling and proposed a
\emph{pseudo-empirical likelihood}.  In brief, their procedure estimates the total of the logarithm of unknown weights in the population using a design unbiased Horvitz-Thompson estimator.  The weights are then determined by maximising this sum under various constraints.  This procedure is based on an estimated likelihood not an observed one.  Rao, Wu and colleagues (notably
\citet{chen_sitter_wu_2002}, \citet{wu_rao_2006},
\citet{rao_wu_2008}, among others) study this method extensively
and apply it to several design based surveys.  The pseudo
empirical likelihood can be re-interpreted as a ``backward''
Kullback-Leibler divergence of the unknown weights from the
sampling weights.  The distribution is specified by the choosing
the weights that minimise this divergence.  \citet{wu_2004}
discuss a similar minimised weighted entropy estimator.  

In this article we develop a framework that results in a procedure which fundamentally differs from the pseudo-empirical likelihood. 
We consider an observed likelihood based on the conditional distribution of the observations given that the individuals were selected in the sample and estimate their distribution in the population using empirical likelihood.  
We are motivated by Pfeffermann and colleagues (e.g. \citet{pfeffermann_1998}, \citet{pfeffermann_sverchkov_1999}, \citet{krieger_pfeffermann_1992}) who used a similar but fully parametric procedure in modelling survey data. 
A previous instance of similar use in a more restrictive parametric set-up occurs in \citet{rao_patil_1978}, where it has been implemented on size-biased sampling.  
We propose a empirical likelihood based semi-parametric approach here. The resulting conditional empirical likelihood is similar to \citet{vardi_1985}.  However, he is motivated solely by the non-parametric estimation of the distribution from multiple samples obtained through different designs.   
The asymptotic properties of this non-parametric estimator have been studied by \citet{gill_vardi_wellner_1988}.  
For complex survey data, a fully non-parametric estimation procedure has been studied by \citet{chambers_dorfman_sverchkov_2003}.  

Use of empirical likelihood in complex data goes back to \citet{qin_1993}, who employed it in a two-sample testing problem, where only one sample was biased by the design.  He showed that under certain conditions the empirical log-likelihood ratio has an asymptotic Chi-squared limit.  A similar approach has been taken by \citet{qin_leung_shao_2002} to analyse data with non-ignorable non-response. 
  \citet{qin_zhang_2007} use an empirical likelihood based method in observational studies where part of the response is missing.  Calibration estimation using a similar empirical likelihood in Poisson sampling has been considered by \citet{kim_2009}.


We start with basic assumptions on the model, design variables and the sampling probabilities, which justifies the analytic form of the likelihood seen in \citet{pfeffermann_1998}.  In this endeavour, the sampling weights are interpreted as random variables depending on all observations of all design variables in the population.  Our framework does not require one to make all the design variables available in the sample. Neither does it assume all observations of the design variables available in the sample are known.
This general framework results in a composite likelihood of the data, which is then converted to an conditional empirical likelihood.  The parameter estimates are obtained by maximising this likelihood under the constraints imposed by the model.  
It is seen that such estimates are strongly consistent and asymptotically normally distributed for the population distribution under usual regular conditions. We also provide estimators of the variances.  
We end by applying our method to estimating electoral success in the $2004$ presidential election in United States of America.

\section{Model and Design Specification}\label{sec:gen}
\subsection{Basic Assumptions and Notations}
We consider a ``superpopulation'' model with response $Y$, a set of auxiliary variables $X=\left\{X^{(1)},X^{(2)},\ldots,X^{(p)}\right\}$ and a set of design variables $D=\left\{D^{(1)},D^{(2)},\ldots,D^{(q)}\right\}$.  The superpopulation is same as the probability space under which the model and the random variables 
are defined.  The set of events also includes all subsets of the set of integers, that is any funnction defined on the all subsets of integers is well defined. These functions are used in defining the sample.        
The population is comprised of $N$ independent and identically distributed draws from the super-population model. We label the elements of the population by $\mc{P}=\{1,2,\ldots,N\}.$ 

A random sample $\mc{S}$ of $n$ observations is drawn from $\mc{P}$ according to a design depending on $D$ and possibly on some unknown parameters (specified in Section \ref{sec:wtdemp}). The available data does not contain all variables in $D$, only a subset $Z=\{Z^{(1)},Z^{(2)},\ldots,Z^{(m)}\}$ is supplied. Let $Z^c=D\setminus Z$.  
Variables in $X$ and $Y$ are not directly involved in the sampling design.  We denote $V=Y\cup X\cup Z$ to be the $m+p+1$ dimensional random vector observed in the data set.  Further, we collect all the explanatory variables in the model in a set $A\subseteq V$.  
Suppose, $D_{\mc{P}}$, $X_{\mc{P}}$, $Y_{\mc{P}}$, $Z_{\mc{P}}$, $Z^c_{\mc{P}}$ denote the vectors and matrices of all observations of the corresponding variables on the population $\mc{P}$.
For $S\subseteq\mc{P}$, $V_{S}$ is the matrix of observations in $S$. $V_{\bar{S}}$ are the observations not in $S$, where $\bar{S} = \mc{P} \setminus S$.  

Primary scientific interest focuses on the relationship between a response $Y$
and the set of explanatory variables $A$.
Examples of such models are generalised linear models (GLM)
\citep{mccneldrbook}.
As an important special case, we consider joint models for $Y$ and $A$, $P_{\theta}(Y,A),$ parametrised by $\theta$.
For example, for GLM $\mu(\theta) = A\theta$. We specify the broader class of
applicable models in Section \ref{sec:modspec} below.
\subsection{Model specification}\label{sec:modspec} 
Suppose $F^0$ is the distribution of $V_1$ in the population with density $dF^0$ w.r.t. a suitable measure. The relationship between the response $Y$ and the set of auxiliary variables $A$ is assumed to be specified by:
\begin{equation}\label{eq:parconstr}
E_{F^0}\left[\psi_{\theta}\left(Y_1,A_1\right)\right]=0.
\end{equation}   
Here $\psi=\left(\psi_1,\ldots,\psi_r\right)$ is a pre-specified function depending only on $Y_1$ and $A_1$ and some unknown parameter $\theta$.  we assume $r$ is at least as big as the dimension of the parameter vector. 
There may be several choices for $\psi$ \citep{Qinlawless1}. For parametric models, such as the GLM
considered in the introduction, 
the corresponding \emph{score functions} $S_{\theta}\left(Y,A\right)$ are
natural choices.  

For simplicity we would assume that the function $\psi$ is well behaved and that the Hessian matrix is continuous near the true value of the parameter (see A.6 in Section \ref{sec:asymp_prop}).  This condition of course can be relaxed.  Medians, quantiles etc. can also be estimated similarly.  

\subsection{Design Specification} 

The sample $\mc{S}$ is a random subset or random multiset (for sampling with replacement) of size $n$ of $\mc{P}$.  If the sample units are drawn according to a design, the sampling mechanism may not be \emph{ignorable}. The observed distribution of $V$ in the sample $\mc{S}$ may be different from its distribution in the population (or the superpopulation) and may depend on the particular sample selected.  
The likelihood of the parameter $\theta$, based on the sample observations of $V$ (ie. $V_{\mc{S}}$) differs from the likelihood based on $V_{\mc{P}}$ (ie. its population observations).  In reality, it is almost impossible to specify the likelihood based on the observations of $V$ from a non-ignorable unequal probability sample.  
The sample at hand would rarely contain all design variables, thus the actual design procedure cannot be determined.  Even if the design mechanism can be determined, incorporating design information in the likelihood is anything but straightforward.
    
\citet{pfeffermann_1998} have introduced a sample likelihood of the parameter based on the population density of $V$ conditional on the event of selection in the sample.  This likelihood demands essentially no design information, 
other than the sampling probabilities of the sampled observations.  It is not however immediately clear that the proposed sample likelihood is meaningful under the population distribution.  Below, we first show the conditions under which the sample likelihood indeed is the true population conditional likelihood, using which interpretable inference about the model parameter can be drawn from the available sample.       
For simplicity, we consider only the subsets of $\mc{P}$ (ie. sampling without replacement) here.  Description for multisets (ie. sampling with replacement) is similar.

For $S\subseteq\mc{P}$ suppose $I_{S}$ is the random indicator function for $S\subseteq\mc{S}$. 
The sampled units are drawn according to a design depending on all observations of the set of design variables in the population (ie. $D_{\mc{P}}$).  With $Pr_{\mc{P}}\left[\cdot\right]$ denoting the probability under the population (and by definition the superpopulation), this implies that, for any $S\subseteq\mc{P}$, the probability of selecting $S$ in the sample is given by:
\begin{equation}
\pi_S=Pr_{\mc{P}}\left[I_S=1\mid D_{\mc{P}}\right].  
\end{equation}

That is, $\pi_S$  represents the (joint) conditional probability for inclusion of subset $S$ in the sample given all observations of all design variables in the population.  
Being conditional probability, for each $S$, $\pi_S$ is a function of $D_{\mc{P}}$, $S$, $n$, $N$ and possibly some other parameters.  That is, $\pi_S$ is a random variable because of $D_{\mc{P}}$.  Since $I_S$ is a binary variable, its distribution is completely specified through $\pi_S$.   
  
We first argue that our definition of $\pi_S$ as a conditional expectation under the population distribution does not conflict with the notion of sampling from a finite population.  To that end, we first assume that:

\begin{assumption}[Conditional independence given the design.]
For all $S\subseteq \mc{P}$, under the population distribution,
$\pi_{S}$ is conditionally independent of $Y_{\mc{P}}$ and $X_{\mc{P}}$ given $D_{\mc{P}}$. That is
\begin{equation}
\cind{\pi_{S}}{\left(Y_{\mc{P}},X_{\mc{P}}\right)}{D_{\mc{P}}}.
~~~~~~~~\mbox{for~all}~ S\subseteq\mc{P}.
\end{equation}  
\end{assumption}

The assumption ensures that $\pi_S$ depends on the $Y_{\mc{P}}$ and $X_{\mc{P}}$ only through the design variables $D_{\mc{P}}$.
Under Assumption $1$ for all $S$, $E_{\mc{P}}\left[\pi_S\mid Y_{\mc{P}},X_{\mc{P}}, D_{\mc{P}}\right]=E_{\mc{P}}\left[\pi_S\mid D_{\mc{P}}\right]=\pi\left(S,D_{\mc{P}}\right)$, for some function $\pi$.  This function may depend on the sample and the population size, but does not depend directly on the distributions of  $X_{\mc{P}}$ and $Y_{\mc{P}}$.  

By construction, $I_S$ is a well-defined random variable under the superpopulation structure. Its distribution is completely determined by $\pi$, which is specified by the practitioner or the sampling procedure used to obtain $\mc{S}$ from $\mc{P}$ and not automatically determined by the super-population model.  
A sampling design can be viewed as a list of $\pi_S$ assigned ideally to every subset $S$ of $\mc{P}$.  That is, once the function $\pi$ is specified, the first-order probabilities of selection for $\{i\}$, $i\in\mc{P}$ are given by $\pi_i=\pi\left(\{i\},D_{\mc{P}}\right), i=1,\ldots, n$. The second-order probabilities are similarly determined by $\pi_{ij}=\pi\left(\{i,j\},D_{\mc{P}}\right)$.  Higher order probabilities can be specified exactly the same way.

In some cases, $\pi_S$ for each $S\subseteq\mc{P}$ is defined in advance and the design is constructed to ensure that the subset $S$ is selected with probability $\pi_S$.  As for example, by definition simple random sampling without replacement procedure ensures that each subset of size $\mid S\mid$ is chosen with pre-specified probability of $\binom{n}{\mid S\mid}/\binom{N}{\mid S\mid}$, for $\mid S\mid\le n$ and zero otherwise.  On the other hand, in some cases a partial list of $\pi_S$ may be available.  Various sampling designs are constructed to match these specified probabilities.  The sampling probabilities of rest of the subsets are then design specified.  
As for example, for probability proportional to size (PPS) sampling the function $\pi$ can be chosen to yield the target first-order probabilities of selection $\pi_{\{i\}}=Z^{(1)}_i/\sum^n_{i=1} Z^{(1)}_i$, for some positive random variable $Z^{(1)}$.  
Several procedures for PPS sampling to sample unit $\{i\}$ with a specified probability $\pi_{\{i\}}$ are known (see \citet{hanif_brewer_1983}, \citet{chaudhuri_voss_1988}, \citet{tille_2006}, among others). Each of these procedures have different higher order selection probabilities.

Once a list of $\pi_S$ is specified for all $S\subseteq\mc{P}$, the physical act of sampling observations from the population $\mc{P}$ to the sample $\mc{S}$ contributes no more to the statistical inference.  That is, once we know the specification of $\pi_S$, the second randomisation and sampling from $\mc{P}$ to $\mc{S}$ can be subsumed under the superpopulation probability structure.  
The concept is akin to the \emph{resampling} procedures popularly used in statistics. The difference as well as the difficulty for non-ignorable sampling is that the population resampled from is unobserved and we can draw only one \emph{weighted} resample.

The concept that the actual sampling procedure can be subsumed in the probability structure of the superpopulation is valid even if Assumption $1$ does not hold. 
From Assumption $1$ and the definition of $\pi_S$ as $E_{\mc{P}}\left[I_S\mid D_{\mc{P}}\right]$ we obtain the following result. 
\begin{lemma}\label{lem:a1}  
Assumption $1$ holds iff $\pi_S=\pi\left(S,D_{\mc{P}}\right)$.
\end{lemma}

Lemma \ref{lem:a1} follows from the definition of conditional independence \citep{lau}.  It further shows that, under Assumption $1$, conditioning on $D_{\mc{P}}$ and the pair $\left(D_{\mc{P}},\pi_S\right)$ is same.  The following relationships can also be obtained from Assumption $1$ and Lemma \ref{lem:a1}.  
 
\begin{lemma}\label{lem:a2} 
Suppose $E_{\mc{P}}\left[\cdot\right]$ denotes the expectation under the population distribution. Under Assumption $1$, for all $S\subseteq\mc{P}$, the following holds:
\begin{enumerate}
\item $E_{\mc{P}}\left[I_S\mid \pi_S\right]=\pi_S$.
\item $\cind{I_S}{D_{\mc{P}}}{\pi_S}$.
\end{enumerate}
\end{lemma}
\citet{pfeffermann_1998} use the relationship in Conclusion $1.$ to justify
their parametric likelihood (see below).  The conditional independence relation in $2.$ is
exactly the ``Condition $1$'' in \citet{sugden_smith_1984}, which implies that
under Assumption $1$ the set of joint probabilities $\{\pi_{S}~:~S\subseteq\mc{P}\}$ contains all design information and given the design the selection procedure (i.e., the actual dependence of
$\pi_{S}$ on $D_{\mc{P}}$) can be ignored for inference.

The assumption that the set of joint selection probabilities contains all information about the
sampling mechanism is natural and facilitates analysis.  In
sample surveys, the probability of selecting an observation
becomes unequal due to clustering, stratification,
post-stratification, attrition, purposive ``oversampling'' and
other non-response adjustments.  In most cases, the published
data does not contain all the design variables, thus the actual
design procedure cannot be determined.  Further, in many cases
large data sets are constructed by merging several available
data sets obtained from different surveys (e.g.
\citet{rendall_et_al_2008, tighe_livert_saxe_2010}).
Typically, each survey is based on different designs dependent on different variables.  A common design for the merged data set is mostly unavailable or may not be easy to specify, but weights from individual surveys can be used to provide information about the underlying designs.

\begin{assumption}[Conditional independence given the sampling probabilities]
For all $S\subseteq\mc{P}$, under the population distribution, $I_S$ is conditionally independent of $X_{\mc{P}}$, $Y_{\mc{P}}$ and  $D_{\mc{P}}$ given $\pi_{S}$. That is,
\begin{equation}
\cind{I_S}{\left(X_{\mc{P}},Y_{\mc{P}},D_{\mc{P}}\right)}{\pi_{S}}
~~~~~~~~\mbox{for~all}~ S\subseteq\mc{P}.
\end{equation}
\end{assumption}

Assumption $2$ implies that inclusion depends on the $X_{\mc{P}},Y_{\mc{P}}$ and $D_{\mc{P}}$ only through the joint inclusion probabilities.
In particular, it implies 
\[
Pr_{\mc{P}}\left[I_S=1\mid X_{\mc{P}},Y_{\mc{P}},D_{\mc{P}},\pi_{S}\right]=Pr_{\mc{P}}\left[I_S=1\mid \pi_{S}\right]=E_{\mc{P}}\left[I_S\mid \pi_S\right]=\pi_S.  
\]
\citet{pfeffermann_1998} make
this assumption without stating it explicitly. 

\begin{lemma}\label{lem:conda2} For all $S\subseteq\mc{P}$,  under Assumptions $1$, Assumption $2$ is equivalent to $\cind{I_S}{D_{\mc{P}}}{\pi_S}$ and $\cind{I_S}{\left(X_{\mc{P}},Y_{\mc{P}}\right)}{D_{\mc{P}}}$.
\end{lemma}
The condition $\cind{I_S}{\left(X_{\mc{P}},Y_{\mc{P}}\right)}{D_{\mc{P}}}$ is the basic design assumption of \citet{scott_1977}.  According to \citet{sugden_smith_1984}, any design which only depends on $D_{\mc{P}}$ should satisfy this condition. 
\begin{lemma}\label{lem:conda3} For all $S\subseteq\mc{P}$, Assumptions $1$ and $2$ imply the following conditional independence relationships.
\begin{enumerate}
\item $\cind{I_S}{V_S}{\pi_S}$,
\item $\cind{\left(I_S,\pi_S\right)}{\left(X_{\mc{P}},Y_{\mc{P}}\right)}{D_{\mc{P}}}$ and
\item $\cind{I_S}{\left(X_S,Y_S\right)}{D_{\mc{P}}}$.
\end{enumerate}
\end{lemma}

The statement $1.$ of Lemma \ref{lem:conda3} implies that $E_{\mc{P}}\left[I_S\mid V_S,\pi_S\right]=E_{\mc{P}}\left[I_S\mid\pi_S\right]=\pi_S$ for all $S\subseteq\mc{P}$. From this, following \citet{pfeffermann_sverchkov_2003} we obtain
\begin{align}
Pr_{\mc{P}}\left[I_S=1\mid V_S\right]&=E_{\mc{P}}\left[I_S\mid V_S\right]=E_{\mc{P}}\left[E_{\mc{P}}\left[I_S\mid V_S,\pi_S\right]\mid V_S \right]=E_{\mc{P}}\left[\pi_S\mid V_S \right],\nonumber\\
Pr_{\mc{P}}\left[I_S=1,V_{S}\right]&=Pr_{\mc{P}}\left[I_S=1\mid V_{S}\right]Pr_{\mc{P}}\left[V_{S}\right]=E_{\mc{P}}\left[\pi_S\mid V_S \right]Pr_{\mc{P}}\left[V_{S}\right].\label{eq:joint1}
\end{align}

Equation \eqref{eq:joint1} shows that under our assumptions the joint probability of selection of a subset $S$ and the measurement of the variable $V$ on $S$ i.e. $V_S$ can be expressed without modelling the design procedure explicitly.  
Expectation of the selection probability conditional on the observation is required.  However, this conditional expectation is a function of the data observed in the sample.  The joint probability does not depend on rest of the unobserved elements in the population.

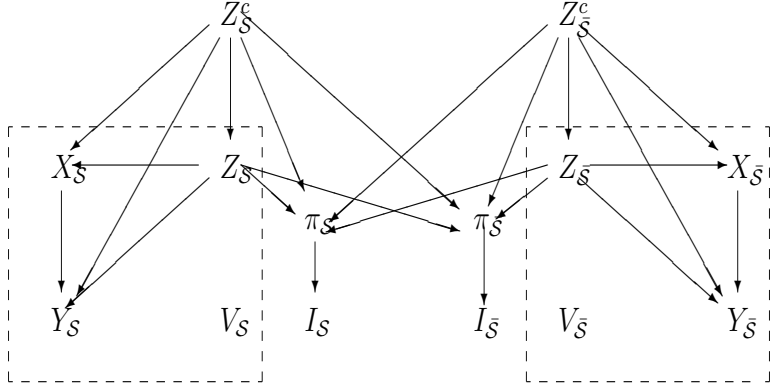
\begin{figure}[t]
\begin{center}
\resizebox{4in}{2in}{\input{graph1.tex}}
\end{center}
\caption{A graphical representation of the assumptions. The directed edges do not necessarily indicate a causal relationship.}
\label{fig:graph}
\end{figure}

A graphical representation of the conditional independencies in Assumptions $1$ and $2$ for $S=\mc{S}$ can be found in Figure \ref{fig:graph}.  In the graph, $\pi_{\mc{S}}$ and $I_{\mc{S}}$ may depend on the whole of $Y_{\mc{P}}$ and $X_{\mc{P}}$ via $Z^c_{\mc{P}}$ which are not available in $\mc{S}$. Thus even though $\pi_{\mc{S}}$ does not depend on $Y_{\mc{S}}$ and $X_{\mc{S}}$ 
directly, in \eqref{eq:joint1}, $E_{\mc{P}}\left[\pi_{\mc{S}}\mid V_{\mc{S}}\right]\ne E_{\mc{P}}\left[\pi_{\mc{S}}\mid Z_{\mc{S}}\right]$ in general. Furthermore, the relation $Y_{S}\ind I_S\mid A_S$ does not hold.  So the design ignorability condition of \citet{pfeffermann_1998} and \citet{pfeffermann_sverchkov_1999} is clearly not satisfied.

Note that our assumptions do not require $I_{\mc{S}}$ to be conditionally independent of $V_{\bar{\mc{S}}}$ given $V_{\mc{S}}$ (see Figure \ref{fig:graph}).  
Thus the observations are not missing at random (in the sense of \citet{little_rubin_book}).  However missing at random is an important special case, because the relationship, $\cind{I_{\mc{S}}}{\left(V_{\bar{\mc{S}}},\pi_{\bar{\mc{S}}}\right)}{\left(V_{\mc{S}},\pi_{\mc{S}}\right)}$ holds.

Finally, we note that, Assumption $2$ is sufficient but not necessary for \eqref{eq:joint1} to hold. One of the conditions $\cind{I_S}{V_S}{\pi_S}$ or $\cind{I_S}{\left(X_S,Y_S\right)}{D_{\mc{P}}}$ would suffice.   We could have alternatively assumed:

{\it Assumption $2^{\prime}$.}  For all $S\subseteq\mc{P}$, under the population distribution, $I_S$ is conditionally independent of $X_{S}$ and $Y_S$ given $D_{\mc{P}}$.  That is,
\begin{equation}
\cind{I_S}{\left(X_{S},Y_S\right)}{D_{\mc{P}}}
~~~~~~~~\mbox{for~all}~ S\subseteq\mc{P}.
\end{equation}

Unlike Assumption $2$, however, Assumption $2^{\prime}$ still allows $I_S$ to be conditionally dependent on $\left(X_{\bar{S}},Y_{\bar{S}}\right)$ given $\pi_S$ without violating Lemma \ref{lem:a2}.  This will happen in very special situations where
typically the information about the design available from $\pi_S$ is incomplete and the design is potentially mis-specified. 
In most cases though Assumption $2$ would be satisfied.

\subsection{A composite likelihood for unequal probability sampling}\label{sec:wtdemp}

Let the $i$th element in $\mc{S}$ be drawn with probability $\pi_i$ (i.e. $\pi_{\{i\}}$), $i=1,2,\ldots,n,$. 
Suppose that $\pi_i$ is positive for $i=1,2,\ldots,N$.

We consider the implication of \eqref{eq:joint1} on each $V_i$ (i.e. $V_{\{i\}}$), $i=1,2,\ldots,n$ selected in the sample.  Let $F^{(i)}_{\mc{S}}$ be the conditional distribution of $V_i$ given $\{i\}\subseteq \mc{S}$, with density $d\pf^{(i)}$.  Using Bayes' rule \citep{pfeffermann_1998}, \eqref{eq:pis} and \eqref{eq:joint1} it follows that:  
\begin{equation}\label{eq:main}
d\pf^{(i)}=\frac{Pr_{\mc{P}}(I_{\{i\}}=1,V_i)}{Pr_{\mc{P}}(I_{\{i\}}=1)}=\frac{E_{\mc{P}}\left[\pi_i\mid V_i\right]dF^0(V_i)}{Pr_{\mc{P}}(I_{\{i\}}=1)},
\end{equation}

where
\begin{equation}\label{eq:D}
Pr_{\mc{P}}(I_{\{i\}}=1)=\int Pr_{\mc{P}}(I_{\{i\}}=1,V_i)dV_i=\int E_{\mc{P}}\left[\pi_i\mid V_i\right]dF^0(V_i)dV_i. 
\end{equation}  

We call the conditional inclusion probability
 $\bp_i{\equiv}E_{\mc{P}}\left[\pi_i\mid V_i\right]$ the {\sl conditional visibility} for the $i$th element in the population and
$\Bp_i\equiv\int \bp_i dF^0(V_i)dV_i = E_{\mc{P}}[\pi_i] = E_{F^0(V_i)}[\bp_i]$
the {\sl visibility factor} for the $i$th element in the population \citep*{rao_patil_1978}.  By substituting these expressions into \eqref{eq:main} we obtain: 
\begin{equation}\label{eq:likhood}
d\pf^{(i)}=\frac{\bp_idF^0(V_i)}{\Bp_i}.
\end{equation} 
To specify $d\pf^{(i)}$ in \eqref{eq:likhood} it is typically necessary 
to model the conditional visibility ($E_{\mc{P}}\left[\pi_i\mid V_i\right]$) 
and the distribution of $V_i$ in the population ($dF^0(V_i)$).
Both of these models may depend on unknown parameters.  We denote the parameter for $dF^{0}(V_i)$ by $\theta$ and that for the model for $E_{\mc{P}}\left[\pi_i\mid V_i\right]$ by $\alpha.$

The composite likelihood for $\alpha$ and $\theta$, using all $V_i$, $i=1,2,\ldots,s$ can now be constructed as: 
\begin{equation}\label{eq:condlik}
L\left(V,\alpha,\theta\right)=\prod^n_{i=1}d\pf^{(i)}.
\end{equation} 
It is similar to the sample likelihood of \citet{pfeffermann_sverchkov_2003}.
Note that \eqref{eq:condlik} does not capture the dependence structure of the $\pf^{(i)}$.  It is a conditional likelihood if the units are drawn independently
of each other, for example, via Poisson sampling.
However, \citet*{pfeffermann_1998} show that for several designs, and under
fairly general conditions, the sampled observations in the conditional
distribution are asymptotically independent as the population 
size $N\to\infty.$ These results suggest that the  \eqref{eq:condlik}
may be a useful surrogate for the conditional likelihood in these
settings. This is also seen in the illustrative example presented in Section \ref{sec:ed} below.

Notice that, \eqref{eq:condlik} is invariant to the scale of $\pi$ and $\bp$,
which can be specified up to an arbitrary positive scaling constant.
We can estimate $\bp_i$ directly from the data if the conditional distribution of $\pi_i$ given $V_i$ in $\mc{S}$ is equal to that in $\mc{P}$. Otherwise, from \citet{pfeffermann_sverchkov_1999}, we obtain $Pr_{\mc{S}}\left(\pi^{-1}_i\mid V_i\right)=\{\pi_i Pr_{\mc{P}}\left(\pi^{-1}_i\mid V_i\right)\}/E_{\mc{P}}\left[\pi_i\mid V_i\right]$. This implies:
\begin{align}
E_{\mc{P}}\left[\pi_i\mid V_i\right]&=\left[E_{\mc{S}}\left(\pi^{-1}_i\mid V_i\right)\right]^{-1}.\label{eq:expi}
\end{align}   

Thus when the sample and population distribution differ a model for $\pi^{-1}_i$ in terms of $V_i$ is sought.  The required population expectation can be estimated from the reciprocal of fitted values of $\pi^{-1}_i$ obtained from the model.  
It is however not clear how to check if the conditional distribution of $\pi$ given $V_i$ in the population and the sample are different.  We reckon that in most cases use of \eqref{eq:expi} would be appropriate. 

\citet{pfeffermann_1998} discuss a class of conjugate parametric models for the distribution of $V_i$
and conditional distribution with $\pi_i$ given $V_i$ such that $d\pf^{(i)}$ is in the same class as $dF^0(V_i)$. This avoids a complicated computation of $\Bp_i$.  However, estimation of $\theta$ is typically complex. The parameters in $d\pf^{(i)}$ usually depends on both $\theta$ and $\alpha$. This is uneconomical since estimates of $\alpha$ usually are not of primary interest.  
Furthermore, \eqref{eq:condlik} may have multiple modes in $\theta$ and $\alpha$ \citep{pfeffermann_sverchkov_1999,pfeffermann_sverchkov_2003}.  Thus $\h{\alpha}$ and $\h{\theta}$ are estimated separately. 

Typically, $\bp_i$ would only depend on a subset of variables in $V$ which may be quite different from $A$. In particular, if the sample $\mc{S}$ was obtained by merging several sub-samples drawn from different designs,  $\bp_i$ depend on the particular sample the $i$th observation belongs to. 
  Such sample indicator variables usually would not be useful in modelling the response.

Parametric estimation of $F^0$ by maximising \eqref{eq:condlik} has been discussed in \citet{rao_patil_1978}.  \citet{vardi_1985, gill_vardi_wellner_1988} consider the corresponding non-parametric likelihood when $\bp=\pi$ and study 
the empirical distribution for biased sampling models in one dimension.

\section{Empirical likelihood to incorporate sampling weights in parameter estimation}\label{sec:emp} 
\subsection{A Conditional Empirical likelihood based formulation}

If $F_0$ is specified by a parametric family $\mc{F}_{\theta}$, a natural way to estimate $\theta$ is to maximise \eqref{eq:condlik} over $\Theta$. 
Direct maximisation of \eqref{eq:condlik} however poses several problems. First of all, analytic expression of $dF^{(i)}_{\mc{S}}$ are available only if one restricts to conjugate families of distributions for $V_i$ and $\bp_i$.   
 Outside this class $\Bp_i$ has to be computed numerically (see \citet{pfeffermann_sverchkov_2003}) which may be time consuming. Furthermore, correct specification of the parametric joint distribution of $V_i$ is difficult in many situations, specially when $V_i$ contains design variables. 
   
An alternative is to use empirical likelihood \citep{owenbook} and estimate $F^0$ non-parametrically from the observed weighted sample and include all the available parametric information in the analysis.

  For the empirical likelihood based formulation the following assumption on $\Bp$ is made. 

\begin{assumption}[Label-independence of the visibility factors]\label{ass:3}
Visibility factors do not depend on the population labels but only on the design variables, i.e.
$\Bp_i=\Bp\left(D_{\mc{P}}\right)\equiv\Bp$.
\end{assumption}

Whether the visibility factors should depend on the individual labels have been discussed in survey sampling literature in the past.  See e.g. \citet*{godambe75,hartleyrao75}. 
Under Assumption \ref{ass:3} each element in the population and sample will have equal visibility factor.  In view of Assumption 1, $\Bp_i=E_{\mc{P}}[\pi_i]$.  Thus Assumption \ref{ass:3} implies $E_{\mc{P}}[\pi_i]$ are all equal in the population.%

We introduce our conditional empirical likelihood.  To that goal, suppose that, for each $F\in\mc{F}$, $w_i=F(\left\{V_i\right\})$
is the weight $F$ assigns on $V_i$ ($w_i=0$ for all $F$ continuous at $V_i$).  Let $\Delta_{n-1}$ denote the $n$ dimensional simplex and for each $\theta\in\Theta$ we define,

\begin{equation}
\mc{W}_{\theta}=\left\{w\in\Delta_{n-1}~:~\sum^n_{i=1}w_i\psi_{\theta}\left(Y_i,A_i\right)=0\right\}\text{ and } \mc{W}=\bigcup_{\theta\in\Theta}\mc{W}_{\theta}.\label{eq:thetaset}
\end{equation}

Under Assumptions 1, 2 and 3, a natural conditional empirical composite likelihood function corresponding to \eqref{eq:condlik} is obtained by substituting $dF^{0}_i=F^{0}(\{V_i\})$ by $w_i$ and $\Bp$ by $\hat{\Bp}=\sum^n_{i=1}\bp_iw_i$. It takes the form:
\begin{equation}\label{eq:lik}
L\left(V,\alpha,\theta\right)=n^n\frac{\prod^n_{i=1}\bp_iw_i}{\left(\sum^n_{i=1}\bp_iw_i\right)^n}.
\end{equation}
The factor $n^n$ is for normalisation.  The log-likelihood is given by:
\begin{equation}\label{eq:logemplik}
L_{CE}(\theta,w,\bp)=n\log(n)+\sum^n_{i=1}\log(\bp_iw_i)-n\log\left(\sum^n_{i=1}\bp_iw_i\right).
\end{equation}
 
In presence of parametric information we estimate the weights $\h{w}_{CE}$ as $\argmax_{w\in\mc{W}}L_{CE}(w,\bp)$.
A constrained estimator $\h{\theta}_{CE}\in\Theta$ can be obtained as \citep{Qinlawless1, chaudhuri_handcock_rendall_2008} 
\begin{equation}\label{eq:theta}
\h{\theta}_{CE}=\argmax_{{\kern-20pt}\theta\in\Theta}\left\{\max_{w\in\mc{W}_{\theta}}\left(L_{CE}(w,\bp)\right)\right\}.
\end{equation}

\citet{kim_2009} considers estimation of population mean under Poisson sampling and uses expression \eqref{eq:logemplik} with $\bp_i$ replaced by $\pi_{i}$.  In the context of two sample testing, \citet{qin_1993} maximises \eqref{eq:logemplik} w.r.t. 
$w_{i}$ and $\Bp$ with the additional constraint $\sum^n_{i=1}w_i\pi_i=\Bp$.  Similar approaches have been taken by \citet{qin_leung_shao_2002, qin_zhang_2007}
to include auxiliary information in the presence of non-ignorable missing observations.

Choice of $\hat{\Bp}$ in the second term of \eqref{eq:logemplik} is crucial.  Our choice involves both $\bp_{i}$ and $w_{i}$.  Use of the sample mean of $\pi$ or $\bp$ would lead to unweighted estimator of the parameters.
  
We follow \citet{pfeffermann_sverchkov_1999,pfeffermann_sverchkov_2003} and estimate $\alpha$ separately from $w$.  In particular, $\h{\alpha}$, the maximum likelihood estimator for $\alpha$
obtained under the model for $E_{\mc{P}}[\pi|V]$ is used to obtain $\bp$. In most cases, our main interest is in finding $\bp$, not $\hat{\alpha}$.

\section{A characterisation of the maximum empirical likelihood estimator}
\subsection{Connection to inverse conditional visibility weighted pseudo-likelihood estimator} 
When the data is collected through a complex design, in order to estimate $\theta$, it is perhaps natural to consider the solutions of the following estimating equations:
\begin{equation}\label{eq:JR}
\sum^n_{i=1}\frac{1}{\bp_i}\psi_{\theta}\left(y_i,a_i\right)=0
\end{equation}

Estimators based on \emph{inverse probability weighted} score functions, as in \eqref{eq:JR}, but with $\bp$ replaced by $\pi$, have been studied in details in the statistics literature. They occur very often in connection with missing data, two-phase designs, etc.  
Even if the conditional visibilities are replaced by fixed sampling probabilities $\pi$,  the left hand side of equation \eqref{eq:JR} is based on a pseudo-likelihood \citep{pfeffermann_sverchkov_2003} and can be seen as a design unbiased Horvitz-Thompson estimator of the total of $\psi_{\theta}\left(y_i,a_i\right)$ in the finite population. \citet{beaumont_2008} uses smoothed version of fixed weights, in practice similar to conditional visibilities in what he regards as a smoothed Horvitz-Thompson estimation.
However, in our case since both $\pi$ and $\bp$ are assumed random, the justification \eqref{eq:JR} as an usual Horvitz-Thompson type estimator is not entirely appropriate.  In this section we show that an alternative explanation using our proposed empirical likelihood based estimator is available.

Suppose $\mc{V}$ is the set of solutions of \eqref{eq:JR} and $\h{\Theta}_{CE}$ is the collection of all $\h{\theta}_{CE}$ in \eqref{eq:theta}. Then we have the following result.

\begin{theorem}\label{thm:eq}
If $\mc{V}$ is non-empty, then $\mc{V}=\h{\Theta}_{CE}$.
\end{theorem}

Our procedure works even when $\mc{V}$ is empty.  Thus the framework introduced above is more general.  Furthermore, it avoids invoking Horvitz-Thompson estimator and provides a better explanation of inverse probability weighted score function based estimators in terms of conditional empirical likelihood.  It is plain to see that our derivation follows naturally from a likelihood framework. 
The resulting log-likelihood is also different from a typical weighted log-likelihood found in the literature.
This can be exploited in Bayesian formulations of related problems specially in small-area estimation and in multi-phase sampling where the design in the later phases depend on the observed variables in the earlier phase (e.g. \citet{breslow_wellner_2006}).
 
If \eqref{eq:JR} has a unique solution, $\h{\theta}_{CE}$ is unique.  Thus if $\psi$ is obtained from a score function corresponding to a generalised linear model, $\h{\theta}_{CE}$ would be unique.  

The following corollary is an easy consequence of Theorem \ref{thm:eq} which provides an estimate of $F^0$.

\begin{corollary}\label{cor:FCE}
When $\mc{V}$ is non-empty, the estimate of $F_0$ obtained by maximising \eqref{eq:logemplik} over $\mc{W}$ is given by:
\begin{equation}
\h{F}_{CE}(C)=\sum^n_{i=1}\frac{(1/\bp_i)}{\sum^n_{i=1}(1/\bp_i)}{\bf 1}_{\{V_i\in C\subseteq \mbb{R}^{m+p+1}\}}.
\end{equation} 
\end{corollary}

\subsection{General Result}
For a given $\theta$, in order to get a constrained estimator of $w$ the objective function is given by:
\begin{equation}\label{eq:objfn}
L\left(w,\lambda_1,\lambda_2\right)=\sum^n_{i=1}\log(w_i)-n\log\left(\sum^n_{i=1}\bp_iw_i\right)-\lambda_1\left(\sum^n_{i=1}w_i-1\right)-n\lambda^T_2\sum^n_{i=1}w_i\psi_i, 
\end{equation} 
where $\psi_i=\psi_{\theta}\left(y_i,a_i\right)$. $\lambda_1$ and $\lambda_2$ are unknown Lagrange multipliers which depend on $\theta$ as well.

By differentiating \eqref{eq:objfn} with respect to $w_i$ and following \citet{owenbook} mutatis mutandis, it follows that $\lambda_1=0$ and with $\kappa=\lambda_2\sum^n_{i=1}\bp_iw_i$ we get:
\begin{equation}\label{eq:hw}
w_i=\frac{\sum^n_{i=1}\bp_iw_i}{n}\frac{1}{\bp_i+\kappa^T\psi_i}.
\end{equation}
 Under our assumptions about $\psi$, (see \citet{Qinlawless1}) $\kappa$ is a continuous differentiable function of $\theta$. It is easily seen that $\kappa$ satisfies the following equation,
\begin{equation}\label{eq:kappa1}
\sum^n_{i=1}w_i\psi_i=\sum^n_{i=1}\frac{\psi_i}{\bp_i+\kappa^T\psi_i}=0.
\end{equation}
The value of $\kappa$ for a given $\theta$ can be determined from \eqref{eq:kappa1}.  Since for each $i$, $w_i\le 1$, from \eqref{eq:hw} for each $i$, $\kappa$ would satisfy the constraint $n\left(\bp_i+\kappa^T\psi_i\right)\ge\sum^n_{i=1}\bp_iw_i$.  Even though our motivation and formulations are completely different, the expression for $w_i$ in \eqref{eq:hw} is similar 
to those obtained by, \citet{kim_2009,berger_delariva_2016,oguz_berger_2016} in their respective settings.  In their formulation however, expressions  with $\bp$ replaced by $\pi$ are obtained.

Now by substituting the value of $w_i$ in the expression of log-likelihood in \eqref{eq:logemplik} we get,
\begin{equation}\label{eq:loglik2}
L_{CE}(\theta,w,\bp)=-\sum^n_{i=1}\log(\bp_i+\kappa^T\psi_i)+\sum^n_{i=1}\log(\bp_i).
\end{equation}
It is evident that \eqref{eq:loglik2} is satisfied by any $\theta$ such that $\mc{W}_{\theta}\ne\emptyset$.  So the derivative, $\partial L_{CE}/\partial \theta$ evaluated at each $\h{\theta}_{CE}$ will be equal to zero. 

The discussion outlined above leads to the following general sets of estimating equations for $\h{\theta}_{CE}$.
 
\begin{theorem}\label{thm:gen}
Under our assumptions, each $\h{\theta}_{CE}\in\h{\Theta}_{CE}$ satisfy the following sets of equations:
\begin{align}
\sum^n_{i=1}\frac{\psi_{\h{\theta}_{CE}}\left(y_i,a_i\right)}{\bp_i+\kappa^T(\h{\theta}_{CE})\psi_{\h{\theta}_{CE}}\left(y_i,a_i\right)}&=0,\label{eq:kappa}\\
\left.\sum^n_{i=1}\frac{\kappa^T(\theta)\psi^{\prime}_{\theta}(y_i,a_i)}{\bp_i+\kappa^T(\theta)\psi_{\theta}\left(y_i,a_i\right)}\right|_{\theta=\h{\theta}_{CE}}&=0\label{eq:deriv}
\end{align}
where $\psi^{\prime}_{\theta^{\prime}}=\partial \psi_{\theta}/\partial \theta|_{\theta=\theta^{\prime}}$.
\end{theorem}

Equations \eqref{eq:kappa} and \eqref{eq:deriv} closely resemble those used by \citet{Qinlawless1} to inspect the asymptotic properties of empirical likelihood based estimators using independent and identically distributed data sets.  These equations are also used in Section \ref{sec:asymp_prop} to derive the asymptotic properties of proposed $\h{\theta}_{CE}$. 

\section{Asymptotic properties and estimators of standard errors}
\subsection{Asymptotic properties}\label{sec:asymp_prop}
We consider the asymptotic properties of $\h{\theta}_{CE}$ under the true population distribution $F^0$ as the population size $N$ grows to infinity. We shall show that these properties of $\h{\theta}_{CE}$ closely resembles those for ordinary empirical likelihood based estimator as described in \citet{Qinlawless1}.  For a formal framework for asymptotic analysis of growing population size we refer to \citet{fuller_2009}.   

Let us denote, $f\left(y,a,\bp,\theta\right)=\psi_{\theta}\left(y,a\right)/\bp$.  Suppose $\theta_0$ be the true value of $\theta$.  We make the following assumptions on $f$ \citep{Qinlawless1,serfling1}:  

\begin{list}{}{}
\item[A.1.] For all $1\le i\le N$, $f\left(y_i,a_i,\bp_i,\theta_0\right)$ are independent and identically distributed random vectors for any $\theta$. 
\item[A.2.] For all $\bp$, $E_{\mc{P}}\left[f(y,a,\bp,\theta_0)\right]=0$.
\item[A.3.] $E_{\mc{P}}\left[f\left(y_1,a_1,\bp_1,\theta_0\right)f\left(y_1,a_1,\bp_1,\theta_0\right)^T\right]$ is positive definite.
\item[A.4.] The Jacobian $\partial f\left(y_1,a_1,\bp_1,\theta\right)/\partial\theta$ is continuous in a neighbourhood of $\theta_0$. Furthermore, in this neighbourhood $||\partial f\left(y_1,a_1,\bp_1,\theta\right)/\partial\theta||$ and $||f\left(y_1,a_1,\bp_1,\theta\right)||^3$ are both bounded by an integrable function $\mc{G}(y,a,\bp)$.
\item[A.5.] $E_{\mc{P}}\left[\partial f\left(y_1,a_1,\bp_1,\theta\right)/\partial\theta|_{\theta=\theta_0}\right]$ has rank $p$.
\item[A.6.] The Hessian matrix $\partial^2 f\left(y_1,a_1,\bp_1,\theta\right)/\partial\theta\partial\theta^T$ is continuous in $\theta$ in a neighbourhood of the true value $\theta_0$ and in this neighbourhood, $||\partial^2 f\left(y_1,a_1,\bp_1,\theta\right)/\partial\theta\partial\theta^T||$ is bounded by some integrable function $\mc{H}(y,a,\bp)$. 
\end{list}
\begin{lemma}\label{lem:exist}
Suppose the assumptions A.1.- A.5. hold. Then, as $N\rightarrow\infty$, under$F_0$, with probability $1$, $L_{CE}$ attains its maximum value at some point $\h{\theta}$ in the interior of the ball $||\theta-\theta_0||\le N^{-1/3}$.  $\h{\theta}$ and $\h{\kappa}=\kappa(\h{\theta})$ satisfy equations \eqref{eq:kappa} and \eqref{eq:deriv}.
\end{lemma}
To prove asymptotic unbiasedness and normality, we first define, $\mG=E_{\mc{P}}\left[\bp^{-1}_1\left.\partial\psi_{\theta}\left(y_1,a_1\right)/\partial\theta\right|_{\theta_0}\right]$ and $\mG^{\star}=E_{\mc{P}}\left[\bp^{-2}_1\psi_{\theta_0}\left(y_1,a_1\right)\psi_{\theta_0}\left(y_1,a_1\right)^T\right]$.

\begin{theorem}\label{thm:clt}
Suppose the assumptions A.1. - A.6. hold. Then equations \eqref{eq:kappa} and \eqref{eq:deriv}, with $n$ replaced by $N$, admits a sequence of solutions $\left(\h{\theta}^{(N)}_{CE},\h{\kappa}^{(N)}\right)$ such that, 
\begin{enumerate}
\item $\left(\h{\theta}^{(N)}_{CE},\h{\kappa}^{(N)}\right)\longrightarrow \left(\theta_0,0\right)$ almost surely, 
\item $\sqrt{N}\left(\h{\theta}^{(N)}_{CE}-\theta_0\right)\Longrightarrow N\left(0,V_{CE}\right)$ in distribution, where
\begin{equation}\label{eq:V}
V_{CE}=\left\{\mG^T(\mG^{\star})^{-1}\mG\right\}^{-1},
\end{equation}
\item $\sqrt{N}\left(\h{\kappa}^{(N)}-0\right)\Longrightarrow N\left(0,V_{\kappa}\right)$ in distribution, where
\begin{equation}\label{eq:VK}
V_{\kappa}=(\mG^{\star})^{-1}\left\{I-\mG V_{CE}\mG^T(\mG^{\star})^{-1}\right\},
\end{equation}
\item $\h{\theta}^{(N)}_{CE}$ and $\h{\kappa}^{(N)}$ are asymptotically independent.
\end{enumerate}
\end{theorem}
\begin{corollary}\label{cor:fr}
If $\mG$ is invertible, $V_{CE}=\mG^{-1}\mG^{\star}(\mG^T)^{-1}$.  Furthermore, $\h{\kappa}$ is asymptotically degenerate at $0$.
\end{corollary}

\subsection{Estimators for finite sample sizes}
For finite sample size $n$, the standard error of $\h{\theta}_{CE}$ can be estimated from a sandwich estimator based on the expression of $V_{CE}$ in \eqref{eq:V}.  To that effect we define $\h{\mG}=\sum^n_{i=1}\h{w}_i\partial\psi_{\theta}\left(y_i,a_i\right)/\partial\theta|_{\theta=\h{\theta}_{CE}}$ and 
$\h{\mG}^{\star}=\sum^n_{i=1}\h{w}^2_i\psi_{\h{\theta}_{CE}}\left(y_i,a_i\right)\psi_{\h{\theta}_{CE}}\left(y_i,a_i\right)^T$.  
The estimated variance of $\h{\theta}_{CE}$ is then given by $\h{V}_{CE}=\left\{\h{\mG}^T(\h{\mG}^{\star})^{-1}\h{\mG}\right\}^{-1}$.  

\subsection{Predictors for finite population}\label{sec:pred}
Suppose that the parameter $\theta$ can be interpreted as a valid summary of the finite population of size $N$ which is the solution of the set of equations $\sum^N_{i=1}\psi_{\theta}\left(y_i,a_i\right)=0$ . Let $\h{\theta}_{\mc{P}}$ is the predicted value of $\theta$ in the population.  Clearly, one choice of $\h{\theta}_{\mc{P}}$ is $\h{\theta}_{CE}$.  

When the sample size is $N$ and each $\pi_i=\bp_i=1$, by Theorem \ref{thm:eq} $\h{w}_i=N^{-1}$ for each $i=1$, $2$, $\ldots$, $N$.  That is $\h{\theta}_{CE}=\h{\theta}_{\mc{P}}$.  Likewise, under the same conditions $\h{V}_{CE}$ becomes

\[
\frac{1}{N}\cdot\left[\frac{1}{N}\sum^N_{i=1}\left.\frac{\partial\psi_{\theta}\left(y_i,a_i\right)}{\partial\theta}\right|_{\h{\theta}_{\mc{P}}}\right]^{-1} \left[\frac{1}{N}\sum^N_{i=1}\psi_{\h{\theta}_{\mc{P}}}\left(y_i,a_i\right)\psi_{\h{\theta}_{\mc{P}}}\left(y_i,a_i\right)^T\right]\left[\frac{1}{N}\sum^N_{i=1}\left.\frac{\partial\psi_{\theta}\left(y_i,a_i\right)}{\partial\theta}\right|_{\h{\theta}_{\mc{P}}}\right]^{-1}.
\]
This is exactly the square of the standard error of $\h{\theta}_{CE}$ obtained from $N$ i.i.d. observations.  This shows that our estimates comply with the assumptions of the superpopulation model.

The prediction variance $V_{\mc{P}}$ for this choice of $\h{\theta}_{\mc{P}}$, is more difficult to estimate.  Such variances should converge to zero as the sample size increases to the population size.  The simple finite population correction leading to $\h{V}_{\mc{P}}\left(\h{\theta}_{\mc{P}}\right)=\left(1-n/N\right)\h{V}_{CE}$ does not seem to be very accurate.  
It should be also noted that, the proposed variance estimate does not include the dependence among the sampled observations or the association between the observations and the inclusion probabilities.  These terms will play a significant role in estimating the prediction variance.

\section{Illustration in the $2004$ U.S. presidential election}\label{sec:ed}
In this section, we illustrate the inference resulting from this framework and compare it to a standard estimator. We consider the county-wise vote counts from the presidential election in United States of America in 2004. The data contains the total votes cast in favour of John Kerry, George W. Bush and Ralph Nader in each of the $N=4600$ counties in the country.  Let $p$ be the proportion of counties where John Kerry had the majority of the cast votes in the election. 
From the data we find that Mr. Kerry won $1507$ counties, which means the true value of the proportion is approximately $0.3276$.  Suppose we want to estimate $p$ from a sample of size $n=40$.  Samples were drawn with probability proportional to the total votes cast in the county for the three candidates.  We use the Till\'{e}, Midzuno and PPS-systematic schemes \citep{cochran_book,tille_2006} to draw our samples.  
The observations from the Till\'{e} scheme  are nearly independent, those from the Midzuno 
scheme observations are slightly more dependent, and those from 
the systematic sample are highly dependent. All samples were drawn using the {\tt sampling} package for {\tt R} \citep{survey,R}.

Suppose $I_i$ is the indicator that Kerry won the $i$th county in the sample.  Thus the constraint imposed by the model on the unknown weights $w$ is given by 
\begin{equation}\label{eq:kcond}
\sum^n_{i=1}w_i\left(I_i-p\right)=0.
\end{equation}
Suppose $c_i$ denote the total votes cast in the $i$th county for the three candidates. The probability of its selection, $\pi_i$, is proportional to $c_i$.  Since $c_i$ is available in the sample, we take $V_i=(I_i,c_i)$ and since $\pi$ is proportional to $c_i$, we get $\bp_i=\pi_i$ for all $i$.

The unique solution of the equation $\sum^n_{i=1}\left(I_i-p\right)\pi^{-1}_i=0$ is 
$\h{p}_{CE}=\sum^n_{i=1}I_i\pi^{-1}_i/\sum^n_{i=1}\pi^{-1}_i$. This coincides with the 
Hajek estimator, usually motivated by design-based considerations. Note that, although the point
estimators coincide, the CE approach provides a different inferential framework.
From Theorem \ref{thm:eq}, $\h{p}_{CE}$ is the unique empirical likelihood based estimate and $\h{w}\propto\pi^{-1}$.  Furthermore, from Theorem \ref{thm:clt} and Corollary \ref{cor:fr}, an estimate of the variance is given by
\begin{equation}\label{eq:v.est}
\h{V}_{CE}=\frac{\sum^n_{i=1}\left(I_i-\h{p}_{CE}\right)^2\pi^{-2}_i}{\left(\sum^n_{i=1}\pi^{-1}_i\right)^2}.
\end{equation}   
We estimate $\h{p}_{\mc{P}}$ and $\h{V}_{\mc{P}}$ as described in Section \ref{sec:pred}.
\begin{figure}[t]
\captionsetup{width=0.9\textwidth}
\subfigure[\label{fig:tille_comp}]{
\resizebox{2.1in}{2.5in}{\includegraphics{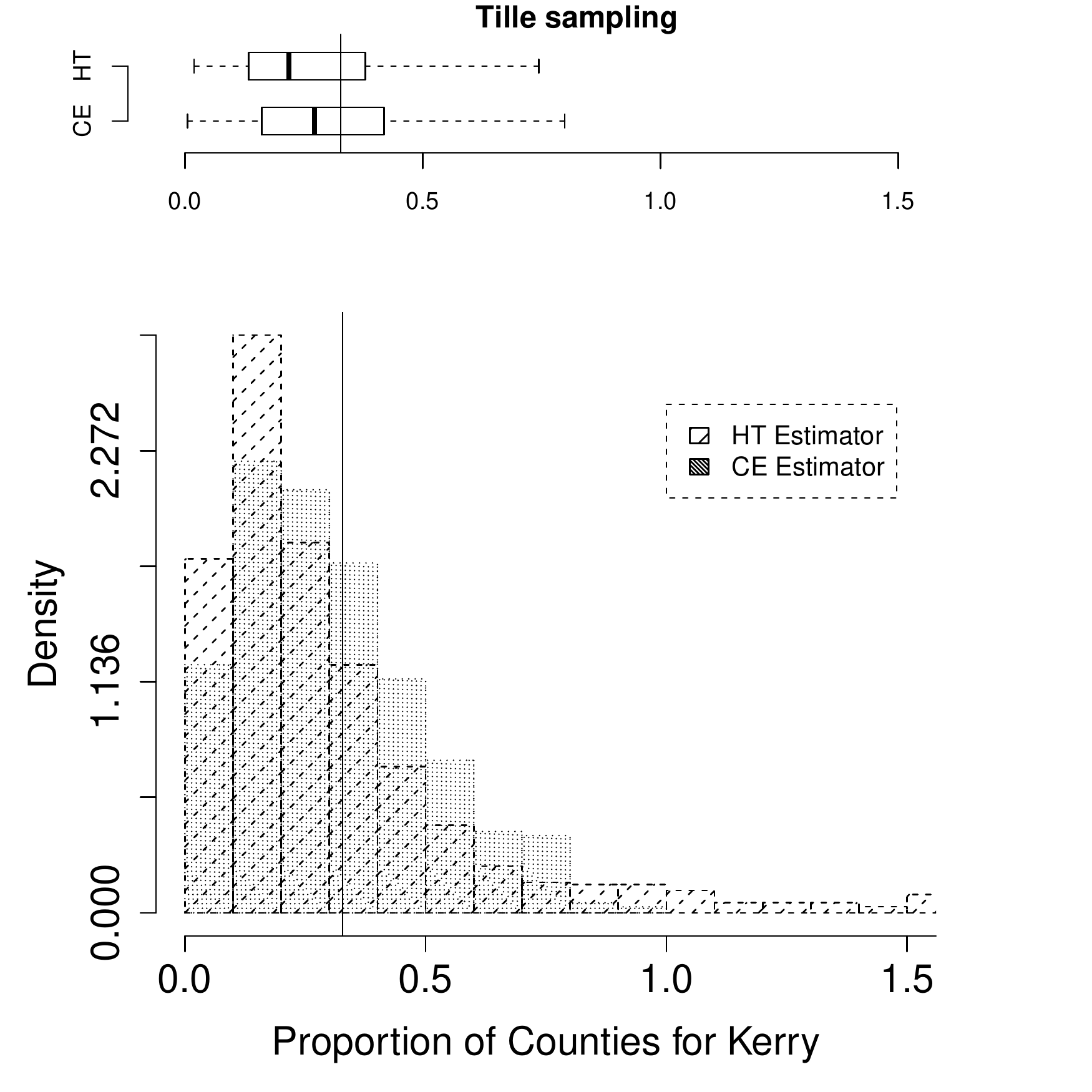}}
}\subfigure[\label{fig:midzuno_comp}]{
\resizebox{2.1in}{2.5in}{\includegraphics{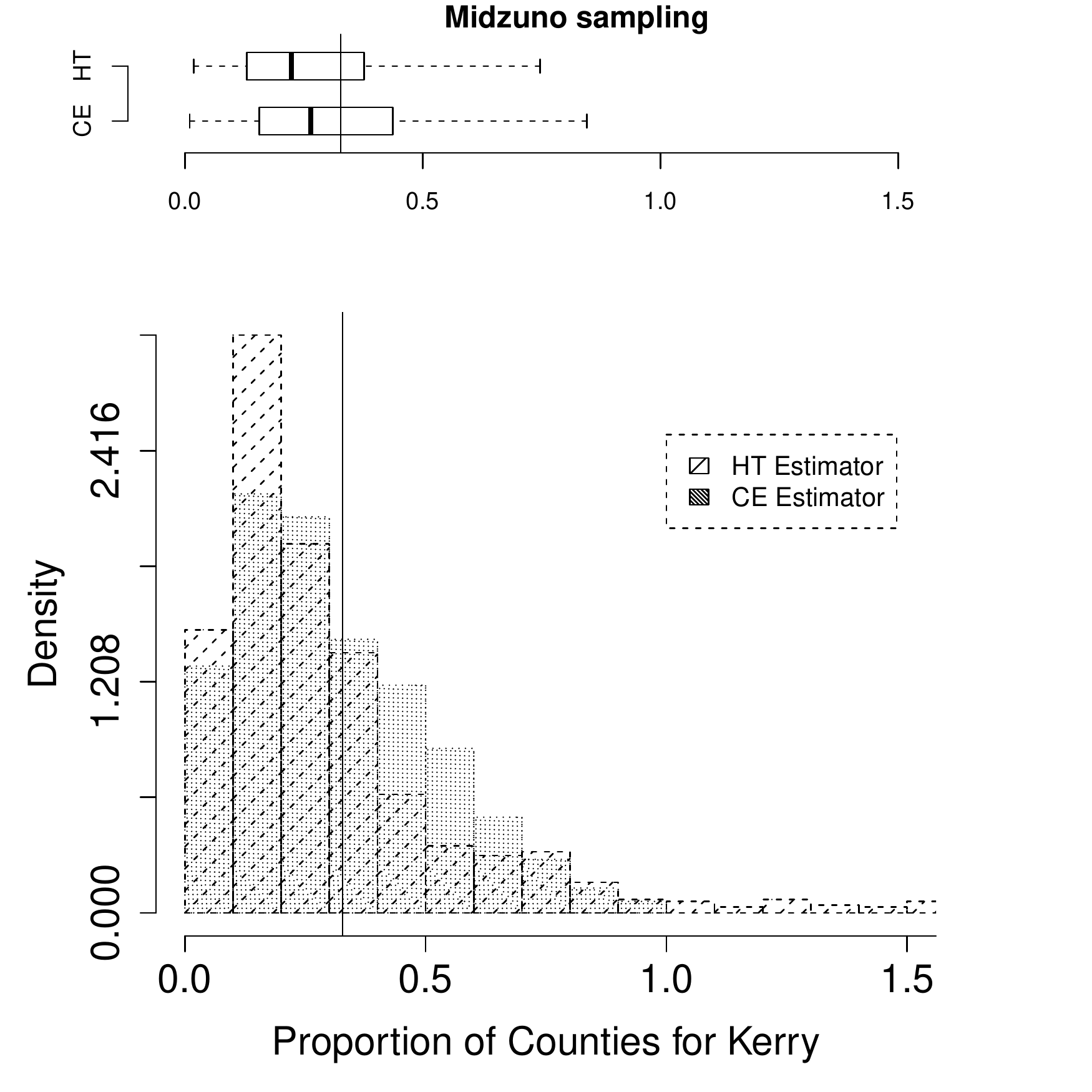}}
}\subfigure[\label{fig:systematic_comp}]{
\resizebox{2.1in}{2.5in}{\includegraphics{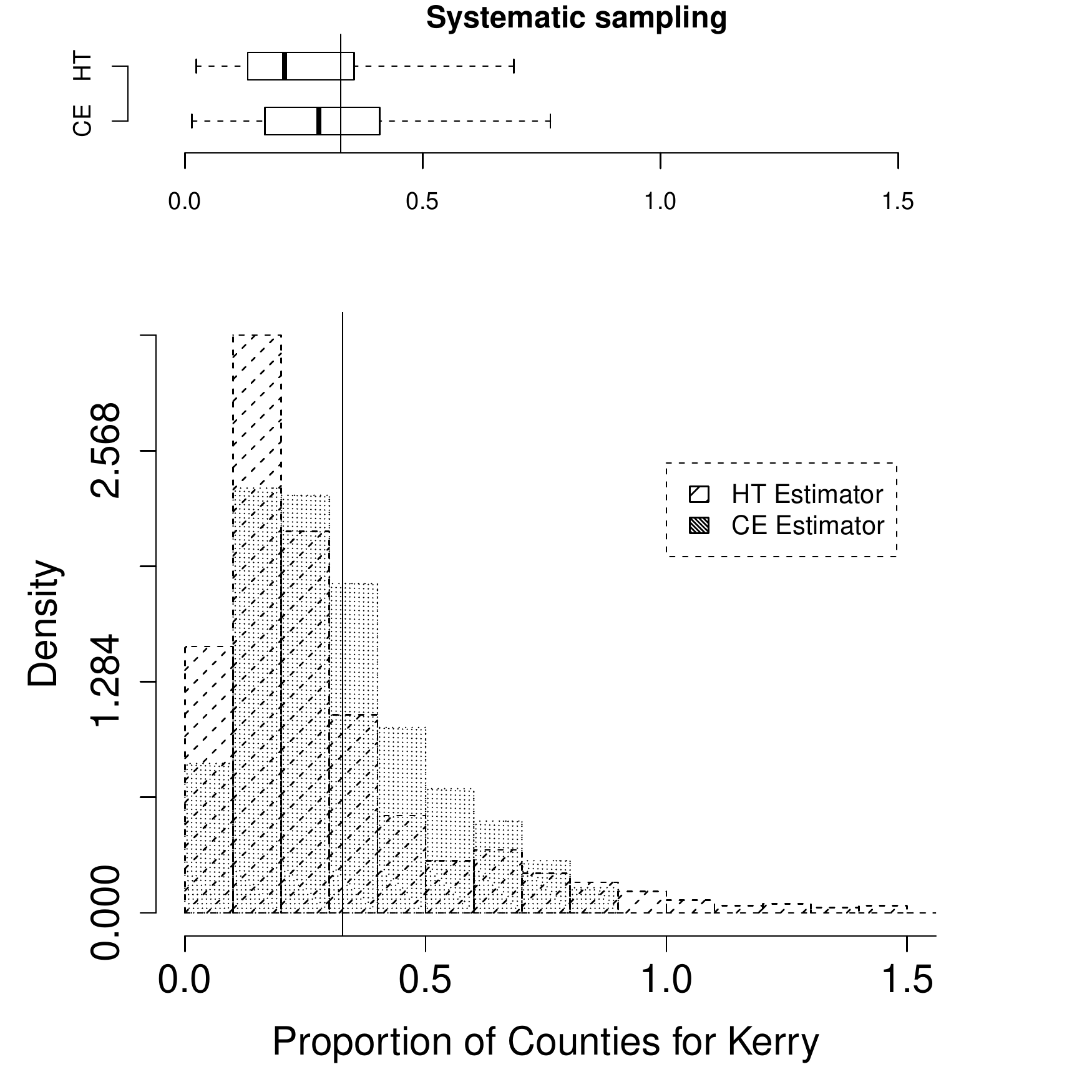}}
}
\caption{Histograms of Horvitz-Thompson and Constrained Empirical Likelihood estimator of the proportion of counties won by John Kerry in the 2004 United States Presidential election.  The plots are based on $1000$ replications each of sample size $40$ collected using \ref{fig:tille_comp} Tille, \ref{fig:midzuno_comp} Midzuno and \ref{fig:systematic_comp} PPS-systematic schemes, respectively. The vertical black line indicates the true value of this proportion in the population, which is $0.3276$.}
\label{fig:comp}
\end{figure}

\begin{table}[ht]
\begin{center}
\caption{Point estimates of the Constrained Empirical Likelihood and the Horvitz-Thompson estimator under three different sampling schemes.  We estimate the proportion of counties won by John Kerry in the $2004$ presidential election. The true value is $0.3276$.}
\begin{tabular}{|l|l|c|}\hline
Sampling&&Mean\\
Scheme&Estimator&Estimate\\\hline
Till\'{e}&Horvitz-Thompson&$0.3376$\\\cline{2-3}
&Conditional EL&$0.3086$\\\hline
Midzuno&Horvitz-Thompson&$0.3205$\\\cline{2-3}
&Conditional EL&$0.3089$\\\hline
Systematic&Horvitz-Thompson&$0.3277$\\\cline{2-3}
&Conditional EL&$0.3111$\\\hline
\end{tabular}

\label{tab:est}
\end{center}
\end{table}

\begin{table}[t]
\caption{Standard errors and coverages for two-sided nominally $95\%$ confidence intervals of the Horvitz-Thompson and the Constrained Empirical Likelihood estimator under three different sampling schemes.}
\begin{tabular}{|l|l|cc|cc|cc|cc|}\hline
Sampling&Estimator&\multicolumn{8}{c|}{Variation}\\\cline{3-10}
scheme&&\multicolumn{2}{c|}{Observed}&\multicolumn{2}{c|}{Conditional EL}&\multicolumn{2}{c|}{Hartley-Rao}&\multicolumn{2}{c|}{Yates-Grundy-Sen}\\\cline{3-10}
&&SE&CI\%&SE&CI\%&SE&CI\%&SE&CI\%\\\hline 
Till\'{e}&HT&$0.383$&$95.0\%$&-&-&$0.386$&$63.1\%$&$0.385$&$63.1\%$\\\cline{2-10}
&CE&$0.188$&$95.1\%$&$0.139$&$72.3\%$&$0.136$&$69.4\%$&$0.136$&$69.4\%$\\\hline
Midzuno&HT&$0.348$&$95.6\%$&-&-&$0.366$&$63.1\%$&$0.342$&$62.0\%$\\\cline{2-10}
&CE&$0.189$&$95.8\%$&$0.136$&$69.4\%$&$0.139$&$71.0\%$&$0.144$&$69.5\%$\\\hline
Systematic&HT&$0.511$&$97.8\%$&-&-&$0.524$&$62.7\%$&NA&NA\\\cline{2-10}
&CE&$0.187$&$95.5\%$&$0.136$&$74.0\%$&$0.137$&$74.5\%$&NA&NA\\\hline
\end{tabular}
\label{tab:res}
\end{table} 

We compare the CE estimator to the Horvitz-Thompson (HT) estimator for a mean, 
given by $\h{p}_{HT}=(4600)^{-1}\sum^n_{i=1}I_i/\pi_i$.

We summarise the results of our study in Figure \ref{fig:comp}, Table \ref{tab:est} and Table \ref{tab:res}.  They are based on an average of $1000$ draws.  Standard errors and the coverage of the two-sided nominally $95\%$ confidence intervals, obtained using Gaussian limits, are also presented.

From the histograms in Figure \ref{fig:comp} it is seen that the Horvitz-Thompson estimator is not always bounded above by one. 
So in some cases the estimates of the proportion is hard to interpret. This is specially true for PPS-systematic sampling where the highest value turns out to be $11.34$ (truncated in Figure \ref{fig:systematic_comp}).  
This estimator turns out to be larger than the one in other two types of sampling as well.  In contrast, the Constrained Empirical Likelihood estimator always varies between zero and one. 
The mean of the Horvitz-Thompson estimator of proportion turns out to be quite close to the correct value for all sampling schemes (see Table \ref{tab:est}).
This is similar to the proposed $\h{p}_{\mc{P}}$. The histograms show that it has a higher modal value than the proposed estimator. However, the skewness of the former is larger.  

In Table \ref{tab:res},
we intend, first, to compare the variation in the Horvitz-Thompson estimator with the Constrained Empirical Likelihood  estimator, and second, to compare the performance of the Constrained Empirical Likelihood  asymptotic variance estimator in \eqref{eq:v.est} with some competing ones.  For comparison we calculate (using the {\tt survey} package in {\tt R} \citep{R}) the Hartley-Rao and Yates-Grundy-Sen estimators of the variance of the Hajek estimator.  Note that, like $\h{V}_{CE}$, the Hartley-Rao estimator only uses the first-order inclusion probabilities. 
The pairwise inclusion probabilities are used in the Yates-Grundy-Sen estimator only.  All three estimators are compared with the observed root mean squared error of $\h{p}_{\mc{P}}$ from $1000$ draws.  This root mean squared error is used as a benchmark in our comparison.  As expected, it has coverage close to the nominal value.    

The root mean squared errors of the Constrained Empirical Likelihood  estimator are insensitive to the choice of sampling procedures.  An exception is the Yates-Grundy-Sen estimator under systematic sampling, where negative estimates of variance are obtained.
We see that all three estimators, that is the proposed $\h{V}_{CE}$, Hartley-Rao and Yates-Grundy-Sen estimators underestimates the observed root mean squared error (when it exists).  This underestimation is expected, since both estimators only approximates the true variance of the Hajek estimator \citep{sarndal_book}. Their performances seems to be comparable.

The Horvitz-Thompson estimator, in general, has higher variance and lower coverage than the proposed estimator.  It is also seen that the average values of Hartley-Rao and Yates-Grundy-Sen estimators are very close to its observed variance (except for systematic sampling).  However, the histograms of these two variance estimators are skewed, which explains the low coverage.             

We intend to make the data and R code available for these procedures on CRAN \citep{R}. The core routines will be compatible with the {\tt survey} package \citep{survey}.

\appendix
\section{Proofs}\label{sec:proof}
In this section we present the proofs of the theorems.\\  
\noindent{\bf Proof of Lemma \ref{lem:a1}}\begin{proof}
Recall that $E_{\mc{P}}\left[I_S\mid D_{\mc{P}}\right]=\pi_S$.  Now Assumption $1$ implies
\[\pi\left(S,D_{\mc{P}}\right)=E_{\mc{P}}\left[\pi_S\mid D_{\mc{P}}\right]=E_{\mc{P}}\left[E_{\mc{P}}\left[I_S\mid D_{\mc{P}}\right]\mid D_{\mc{P}}\right]=E_{\mc{P}}\left[I_S\mid D_{\mc{P}}\right]=\pi_S.\] The other side is immediate.
\end{proof}

\noindent{\bf Proof of Lemma \ref{lem:a2}}\begin{proof}
\begin{enumerate}
\item Using $\pi_S=\pi\left(S,D_{\mc{P}}\right)$,  $E_{\mc{P}}\left[I_S\mid \pi_S,D_{\mc{P}}\right]=E_{\mc{P}}\left[I_S\mid D_{\mc{P}}\right]=\pi_S$.  This means 
\begin{equation}\label{eq:pis}
E_{\mc{P}}\left[I_S\mid \pi_S\right]=E_{\mc{P}}\left[E_{\mc{P}}\left[I_S\mid \pi_S,D_{\mc{P}}\right]\mid \pi_S\right]=E_{\mc{P}}\left[E_{\mc{P}}\left[I_S\mid D_{\mc{P}}\right]\mid \pi_S\right]=\pi_S.\end{equation}

\item Clearly  $Pr_{\mc{P}}\left[I_S=1\mid \pi_S,D_{\mc{P}}\right]=\pi_S$.  Since $I_S$ is binary, its conditional distribution given $\pi_S$ and $D_{\mc{P}}$ is a function of $\pi_S$ only.  So from the definition of conditional independence \citep{lau} the result follows.
\end{enumerate}
\end{proof}

\noindent{\bf Proof of Lemma \ref{lem:conda2}}\begin{proof} From \citet[page 29]{lau} it can be shown that, $\cind{I_S}{\left(X_{\mc{P}}, X_{\mc{P}}, D_{\mc{P}}\right)}{\pi_S}$ is equivalent to $\cind{I_S}{D_{\mc{P}}}{\pi_S}$ and 
$\cind{I_S}{\left(X_{\mc{P}}, X_{\mc{P}}\right)}{\left(\pi_S,D_{\mc{P}}\right)}$.  From Lemma \ref{lem:a1}, under Assumption $1$, the second conditional independence relationship is equivalent to $\cind{I_S}{\left(X_{\mc{P}}, X_{\mc{P}}\right)}{D_{\mc{P}}}$.   
\end{proof}

\noindent{\bf Proof of Lemma \ref{lem:conda3}}\begin{proof}  The proofs follow from \citet[page 29]{lau}. We only present a sketches.
\begin{enumerate}
\item Follows from Assumption $2$.  \\
\item From Assumption $2$, it follows that $\cind{I_S}{\left(X_{\mc{P}}, X_{\mc{P}}\right)}{\left(\pi_S,D_{\mc{P}}\right)}$ holds.  This together with Assumption $1$ completes the proof.\\
\item This statement follows from $2.$ above.
\end{enumerate}
\end{proof}

\noindent{\bf Proof of Theorem \ref{thm:eq}}\begin{proof}  Since the geometric mean is bounded by the arithmetic mean.  Clearly the relationship $\left(\prod^n_{i=1}n\bp_iw_i\right)/\left(\sum^n_{i=1}\bp_iw_i\right)^n\le 1$ holds.  The equality holds iff $w_i\propto\bp_i^{-1}$ for each $i=1,2,\ldots,n$.

Now for any $\h{\theta}\in\mc{V}$, $\sum^n_{i=1}\psi_{\h{\theta}}\left(y_i,a_i\right)/\bp_i=0$. Thus $\h{w}_i(\h{\theta})=\bp^{-1}_i/\sum^n_{i=1}\bp^{-1}_i$, $i=1,2,\ldots,n$ satisfy the constraints. Furthermore, $\left\{\prod^n_{i=1}n\bp_i\h{w}_i(\h{\theta})\right\}/\left\{\sum^n_{i=1}\bp_i\h{w}_i(\h{\theta})\right\}^n=1$.  Thus $\h{\theta}\in\h{\Theta}$ and $\mc{V}\subseteq \h{\Theta}$.

Now let $\theta\in\h{\Theta}\setminus\mc{V}$.  For each fixed $\theta$, $\mc{W}_{\theta}$ is a convex set and the log-likelihood \eqref{eq:logemplik} is a concave function.  Thus the latter has a unique maxima $\h{w}$.  
Since $\sum^n_{i=1}\psi_{\theta}\left(y_i,a_i\right)/\bp_i\ne0$, $\h{w}_i(\theta)\ne\bp^{-1}_i/\sum^n_{i=1}\bp^{-1}_i$, for at least one $i$.  Thus $\left\{\prod^n_{i=1}n\bp_i\h{w}_i(\theta)\right\}/\left\{\sum^n_{i=1}\bp_i\h{w}_i(\theta)\right\}^n<1$.  
However, since $\mc{V}\ne\emptyset$, there is a $\theta^{\star}\in\mc{V}$ such that $\left\{\prod^n_{i=1}n\bp_i\h{w}_i(\theta^{\star})\right\}/\left\{\sum^n_{i=1}\bp_i\h{w}_i(\theta^{\star})\right\}^n>\left\{\prod^n_{i=1}n\bp_i\h{w}_i(\theta)\right\}/\left\{\sum^n_{i=1}\bp_i\h{w}_i(\theta)\right\}^n$.  This implies $\theta\not\in\h{\Theta}$ and $\h{\Theta}\setminus\mc{V}=\emptyset$.
\end{proof}

\noindent{\bf Proof of Corollary \ref{cor:FCE}}\begin{proof}
Evident from the definition of $\h{F}_{CE}$ and Theorem \ref{thm:eq} above.
\end{proof}

\noindent{\bf Proof of Lemma \ref{lem:exist}}\begin{proof}
Note that, for any $\theta$, equation \eqref{eq:kappa} can be re-expressed as,
\begin{equation}
\sum^n_{i=1}\frac{\psi_{\theta}(y_i,a_i)/\bp_i}{1+\kappa(\theta)\psi_{\theta}(y_i,a_i)/\bp_i}=0.
\end{equation}  
Now the rest of the proof follows from \citet{owenbook} and \citet[Lemma 1.]{Qinlawless1}, mutatis mutandis. 
\end{proof}
\noindent{\bf Proof of Theorem \ref{thm:clt}}\begin{proof}
The proof is very close to that of \citet[Theorem 1]{Qinlawless1}.  We expand \eqref{eq:kappa} and \eqref{eq:deriv} around $(\theta_0,0)$.  After some algebra this yields
\[\begin{pmatrix}
&\h{\kappa}^{(N)}\\
&\h{\theta}^{(N)}_{CE}-\theta_0
\end{pmatrix}=J^{-1}_N\begin{pmatrix}
&-\frac{1}{N}\sum^N_{i=1}f(y_i,a_i,\bp_i,\theta_0)+o_p(\delta_N)\\
&o_p(\delta_N)
\end{pmatrix},
\]  
where 
\[J_N=\begin{pmatrix}
-\frac{1}{N}\sum^N_{i=1}\frac{1}{\bp^2_i}\psi_{\theta_0}(y_i,a_i)\psi_{\theta_0}(y_i,a_i)^T&\frac{1}{N}\sum^N_{i=1}\frac{1}{\bp_i}\left.\frac{\partial \psi_{\theta}(y_i,a_i)}{\partial \theta}\right|_{\theta=\theta_0}\\
\frac{1}{N}\sum^N_{i=1}\frac{1}{\bp_i}\left.\left(\frac{\partial \psi_{\theta}(y_i,a_i)}{\partial \theta}\right)^T\right|_{\theta=\theta_0}&0
\end{pmatrix}\longrightarrow\begin{pmatrix}
-\mG^{\star}&\mG\\
\mG^T&0
\end{pmatrix}\text{ a.s.}
\]
and $\delta_N=||\h{\theta}^{(N)}_{CE}-\theta_0||+||\h{\kappa}^{(N)}||=O_p\left(N^{-1/2}\right)$.  
The proof follows since $\sum^N_{i=1}f(y_i,a_i,\bp_i,\theta_0)/N=O_p(N^{-1/2})$.  
\end{proof}

\noindent{\bf Proof of Corollary \ref{cor:fr}}\begin{proof}
The first part is trivial. For the second part note that if $\mG$ is invertible, $V_{\kappa}=0$.
\end{proof}

\bibliographystyle{chicago}

\input{chaudhuriHandcock.bbl}
\end{document}

%% file: graph1.tex
\setlength{\unitlength}{3947sp}%
\begingroup\makeatletter\ifx\SetFigFont\undefined%
\gdef\SetFigFont#1#2#3#4#5{%
  \reset@font\fontsize{#1}{#2pt}%
  \fontfamily{#3}\fontseries{#4}\fontshape{#5}%
  \selectfont}%
\fi\makeatother\endgroup%
\begin{picture}(5424,2247)(2089,-2173)
\put(6001,-61){\makebox(0,0)[lb]{\smash{{\SetFigFont{12}{14.4}{\rmdefault}{\mddefault}{\updefault}{$Z^c_{\bar{\mc{S}}}$}%
}}}}
\thinlines
{\put(3601,-136){\vector(-2,-3){1015.385}}
}%
{\put(2476,-1036){\vector( 0,-1){600}}
}%
{\put(3451,-886){\vector(-1, 0){900}}
}%
{\put(3526,-961){\vector(-4,-3){1020}}
}%
{\put(3751,-136){\vector( 1,-2){450}}
}%
{\put(3751,-886){\vector( 4,-3){384}}
}%
{\put(5926,-61){\vector(-4,-3){1548}}
}%
{\put(5926,-886){\vector(-4,-1){1570.588}}
}%
{ \put(3751,-886){\vector( 4,-1){1552.941}}
}%
{ \put(4276,-1336){\vector( 0,-1){300}}
}%
{ \put(6076,-136){\vector( 0,-1){600}}
}%
{ \put(3676,-136){\vector( 0,-1){600}}
}%
{ \put(6151,-61){\vector( 4,-3){984}}
}%
{ \put(6226,-886){\vector( 1, 0){975}}
}%
{ \put(6151,-136){\vector( 2,-3){1015.385}}
}%
{ \put(7276,-1036){\vector( 0,-1){600}}
}%
{ \put(6151,-961){\vector( 4,-3){984}}
}%
{ \put(6001,-136){\vector(-1,-2){495}}
}%
{ \put(5926,-961){\vector(-3,-2){363.462}}
}%
{ \put(5476,-1261){\vector( 0,-1){450}}
}%
{ \put(3751, 14){\vector( 4,-3){1548}}
}%
{ \put(2101,-2161){\dashbox{60}(1800,1500){}}
}%
{ \put(5776,-2161){\dashbox{60}(1725,1500){}}
}%
\put(2401,-961){\makebox(0,0)[lb]{\smash{{\SetFigFont{12}{14.4}{\rmdefault}{\mddefault}{\updefault}{$X_{\mc{S}}$}%
}}}}
\put(3601,-961){\makebox(0,0)[lb]{\smash{{\SetFigFont{12}{14.4}{\rmdefault}{\mddefault}{\updefault}{$Z_{\mc{S}}$}%
}}}}
\put(2401,-1861){\makebox(0,0)[lb]{\smash{{\SetFigFont{12}{14.4}{\rmdefault}{\mddefault}{\updefault}{$Y_{\mc{S}}$}%
}}}}
\put(3601,-1861){\makebox(0,0)[lb]{\smash{{\SetFigFont{12}{14.4}{\rmdefault}{\mddefault}{\updefault}{$V_{\mc{S}}$}%
}}}}
\put(4201,-1261){\makebox(0,0)[lb]{\smash{{\SetFigFont{12}{14.4}{\rmdefault}{\mddefault}{\updefault}{$\pi_{\mc{S}}$}%
}}}}
\put(4201,-1861){\makebox(0,0)[lb]{\smash{{\SetFigFont{12}{14.4}{\rmdefault}{\mddefault}{\updefault}{$I_{\mc{S}}$}%
}}}}
\put(5401,-1261){\makebox(0,0)[lb]{\smash{{\SetFigFont{12}{14.4}{\rmdefault}{\mddefault}{\updefault}{$\pi_{\bar{\mc{S}}}$}%
}}}}
\put(5401,-1861){\makebox(0,0)[lb]{\smash{{\SetFigFont{12}{14.4}{\rmdefault}{\mddefault}{\updefault}{$I_{\bar{\mc{S}}}$}%
}}}}
\put(6001,-961){\makebox(0,0)[lb]{\smash{{\SetFigFont{12}{14.4}{\rmdefault}{\mddefault}{\updefault}{$Z_{\bar{\mc{S}}}$}%
}}}}
\put(7201,-961){\makebox(0,0)[lb]{\smash{{\SetFigFont{12}{14.4}{\rmdefault}{\mddefault}{\updefault}{$X_{\bar{\mc{S}}}$}%
}}}}
\put(7201,-1861){\makebox(0,0)[lb]{\smash{{\SetFigFont{12}{14.4}{\rmdefault}{\mddefault}{\updefault}{$Y_{\bar{\mc{S}}}$}%
}}}}
\put(6001,-1861){\makebox(0,0)[lb]{\smash{{\SetFigFont{12}{14.4}{\rmdefault}{\mddefault}{\updefault}{$V_{\bar{\mc{S}}}$}%
}}}}
\put(3601,-61){\makebox(0,0)[lb]{\smash{{\SetFigFont{12}{14.4}{\rmdefault}{\mddefault}{\updefault}{$Z^c_{{\mc{S}}}$}%
}}}}
{ \put(3526,-61){\vector(-4,-3){984}}
}%
\end{picture}%